\documentclass{article}

\usepackage[preprint]{main}
\makeatletter
\renewcommand{\@notice}{}
\makeatother
\usepackage[utf8]{inputenc} 
\usepackage[T1]{fontenc}    
\usepackage{hyperref}       
\usepackage{url}            
\usepackage{booktabs}       
\usepackage{amsmath}        
\usepackage{amsfonts}       
\IfFileExists{nicefrac.sty}{\usepackage{nicefrac}}{}       
\IfFileExists{microtype.sty}{\usepackage{microtype}}{}     
\usepackage{xcolor}         
\usepackage{graphicx}
\usepackage{etoolbox}
\setcitestyle{numbers,square,comma}
\makeatletter
\pretocmd{\NAT@citex}{\leavevmode}{}{}
\AtBeginDocument{%
  \let\bsim@natlinkstart\hyper@natlinkstart
  \let\bsim@natlinkend\hyper@natlinkend
  \renewcommand\hyper@natlinkstart[1]{\leavevmode\hbox\bgroup\bsim@natlinkstart{#1}}%
  \renewcommand\hyper@natlinkend{\bsim@natlinkend\egroup}%
}
\makeatother
\title{A Differentiable Surrogate for Loss-Dominated Ion Beam Transport}
\workshoptitle{Sim2Science}

\begin{document}
\author{%
  J.~M. Munoz\thanks{ \texttt{josemm@mit.edu}, \texttt{rgarciar@mit.edu}} \quad
  L. Ruiz-Arango \quad D. Gonzalez-Acevedo \\[2pt]
  {\bfseries F.~C. Pastrana-Cruz \quad J. Hacias \quad R.~F. Garcia~Ruiz\footnotemark[1]} \\[4pt]
  Laboratory for Nuclear Science, Massachusetts Institute of Technology, \\
  Cambridge, Massachusetts, USA
}

\maketitle

\begin{abstract}

Radioactive nuclei with extreme proton-to-neutron ratios
provide unique probes of nuclear structure, fundamental
symmetries, astrophysical processes, and applications such
as medicine.
However, their short half-lives and low production rates mean
that experiments must work with few desired ions, often amid
contaminant backgrounds exceeding the signal by six orders
of magnitude.
Efficiently isolating,
transporting, and delivering these ions through multi-stage
apparatus is therefore a central challenge.
We introduce a differentiable probabilistic surrogate for ion beam
transport in regimes dominated by particle loss.
The model jointly captures the survival
probability of ions along the beamline and the evolving phase-space distribution
of the surviving particles by combining discrete-time survival analysis with
conditional normalizing flows. Applied to a high-dimensional electrostatic
beamline, the resulting likelihood enables gradient-based optimization of
operating voltages and inference of beam properties from sparse detector
observations.
Once trained, the surrogate provides millisecond-scale differentiable predictions, transforming particle-tracking simulations with hard losses into an inference-ready model for experimental design, beamline optimization, and future digital twins of experiments with rare nuclei.
\end{abstract}

\section*{Introduction}

Experimental physics requires accurate control of apparatus governed by coupled operating parameters, imperfect environmental conditions, and expensive simulations. This is especially challenging for experiments with short-lived radioactive nuclei, which are produced at low rates amid large contaminant backgrounds. Radioactive nuclei with extreme
proton-to-neutron ratios provide unique probes of nuclear structure,
fundamental symmetries, astrophysical processes, and applications in medicine,
but experiments may have access to only a handful of desired ions~\citep{udrescu2026}.

These experiments must bridge a severe energy mismatch from nuclear reactions above MeV to sub-meV energy scales in precision studies, which require slow, cold ion beams~\citep{yang2023laser,arrowsmith2023}. The ions must be stopped, separated from contaminants, cooled, and transported between gas cells, guides, and ion traps, while dissipating energy through several controlled buffer-gas stages, while some regions maintain ultra-high vacuum. Transporting the desired ions over meters through millimeters-wide conductance-limiting openings makes particle loss a challenge. Increasing transmission alone, however, is insufficient as the surviving beam must have the shape and energy distribution required downstream, while all elements remain within operating limits~\citep{wiedemann2015,newhauser2015,chen2016,cruz2026crema}.

Numerical particle-tracking simulations are widely used to design and optimize
such systems~\citep{kalvas2010ibsimu,adelmann2019opal}. They
encode detailed field maps and beamline geometries, but repeated evaluations make design, optimization, uncertainty quantification, and inference
computationally prohibitive.
An optimization effort of the setup configuration commonly requires order \(10^8\) ion trajectories, which might lead to weeks of compute. The difficulty is amplified in loss-dominated transport with non-differentiable changes in transmission and leaves only sparse information
about the particles that survive to downstream detectors.

Learning a surrogate of a physical simulator has become a common
strategy across science. Once trained on simulated data, a surrogate provides
a fast approximation to the map from operating conditions to physical
observables, and can expose differentiable structure for optimization and inference~\citep{10.3389/fdata.2022.787421,pipi2026,munoz2026,cuoco2025applications}.
In accelerator physics, surrogate models have similarly accelerated design,
optimization, and online tuning~\citep{edelen2018opportunities,edelen2020surrogate,duris2020bayesian},
while generative models have reconstructed beam phase space and generated particle ensembles for fast detection
and beam simulation~\citep{gpsr, diffusioncloud,caloGAN2018,hashemi2024deepgen}. However, particle transport with substantial beam loss is still a challenge in the field.

Here, we introduce a differentiable probabilistic surrogate for loss-dominated
ion beam transport, as commonly encountered in rare-isotope experiments. We
factorize the mixed simulator response into the probability of survival to each
detector plane and the conditional phase-space distribution of surviving ions.
Combining discrete-time survival analysis~\citep{prentice1978grouped,tutz2016modeling}
with a conditional normalizing flow~\citep{flow} yields a unified, tractable
likelihood for particle counts and phase-space observations. We demonstrate the
approach on SIMION simulations~\citep{dahl2000simion} of a molecular-ion beamline for injection into a cryogenic buffer-gas
cell. The surrogate predicts transmission and six-dimensional phase space at
the simulated detector planes, enabling gradient-based optimization under
distributional constraints and inference at unobserved locations. It thereby transforms particle-tracking simulations with hard losses and
non-differentiable transmission into inference-ready surrogates for
experimental design and beamline optimization, providing a foundation for
future digital twins of precision experiments with rare ions.

\begin{figure}[t]
\centering
\vspace{-4pt}
\includegraphics[width=0.8\textwidth]{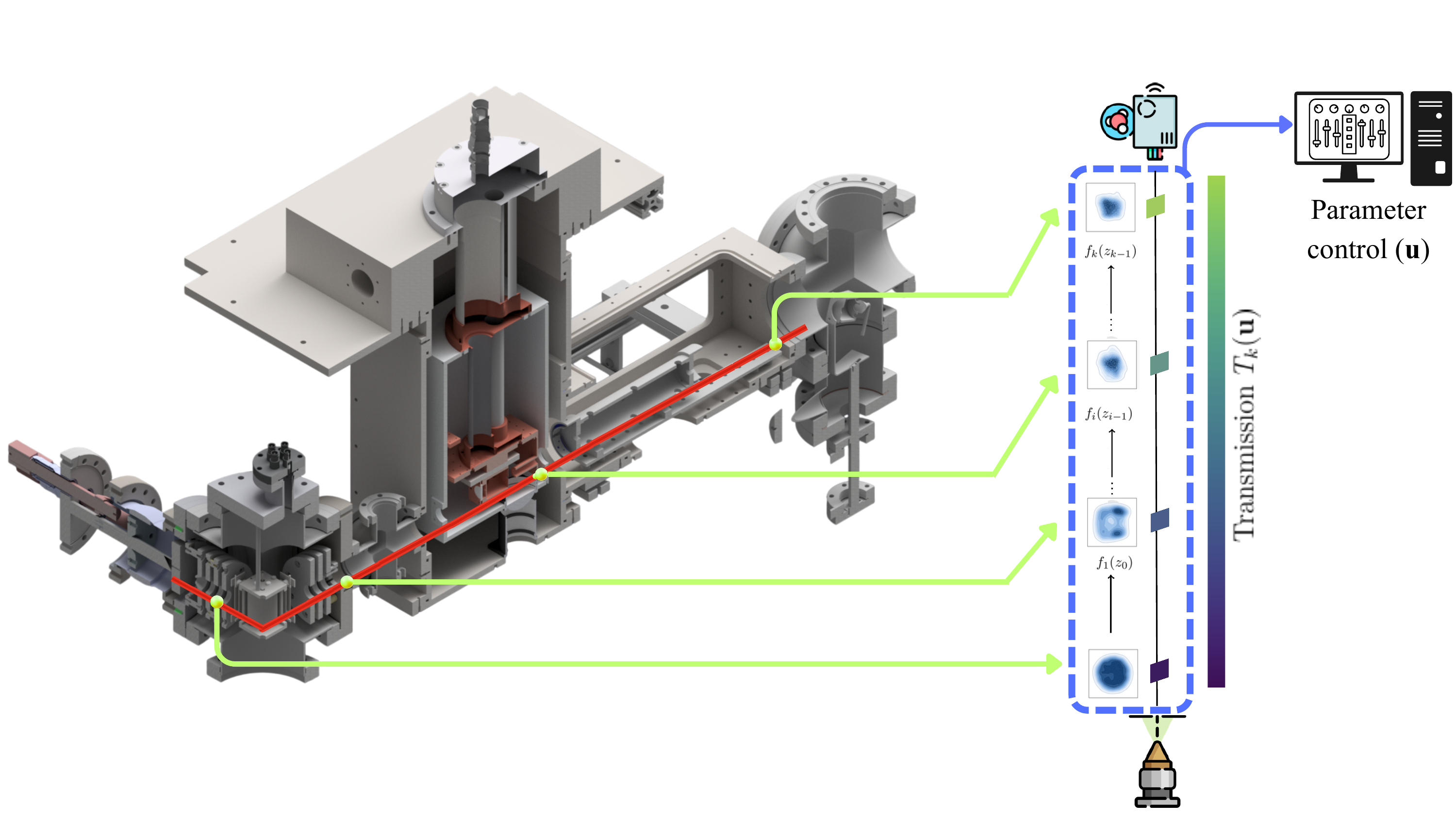}
  \vspace{-4pt}
  \caption{\textbf{The surrogate.}
  Illustration of the estimation of the ion phase space at any detector plane conditioned
  on the voltage configuration going into the setup, alongside the mapped experimental design.
  The drawn plane-to-plane maps are schematic.}
  \label{fig:model}
  \vspace{-4pt}
\end{figure}
\section*{Methods}

To construct a physics surrogate of the experiment, we first generate a
particle-tracking dataset with SIMION~\citep{dahl2000simion}. For each
electrode-voltage configuration, SIMION solves the electrostatic boundary-value
problem on a $1\,\mathrm{mm}$ grid, yielding the electric potential and the
corresponding electric field of the beamline geometry. Ions are then launched
from a specified initial phase-space distribution and propagated through these
fields by numerical integration of their equations of motion. At each detector
plane, we record both the number of ions that survive to that location and the
six-dimensional phase-space coordinates of the surviving particles.
The experimental setup is controlled by a set of voltages
$\mathbf{u}\in\mathbb{R}^{22}$, and its response is observed at ten detector
planes located at arbitrary arc lengths $s_1<\dots<s_{10}$ along the beam axis, as shown in
Figure~\ref{fig:model}.

We generated 65k electrode voltage configurations via Latin hypercube sampling, complemented by configurations visited during a black-box optimization on the transmission to the last detector ($T_{10}$) and stochastic perturbations around the found maximum.
For every voltage configuration, each detector plane records
the six-dimensional single-ion phase space $\mathbf{x}\in\mathbb{R}^6$, of positions, transverse angles, and kinetic energies for the $10^4$ simulated ions. Appendix~\ref{app:simion} provides more details.

We model ion transport with particle loss using a multiplicative ansatz over
the surviving mass and beam shape.
For $N$ ions launched, the expected number reaching plane $k$ within a
phase-space element $d\mathbf{x}$ is
\begin{equation}
  \widehat{\mu}_k(d\mathbf{x}\mid\mathbf{u})
  \;=\;
  N\, T_k(\mathbf{u})\,
  p_\theta(\mathbf{x}\mid\mathbf{u}, s_k)\, d\mathbf{x},
  \label{eq:ansatz}
\end{equation}
with the transmission $T_k(\mathbf{u})\in[0,1]$ setting the surviving mass and the
density $p_\theta(\mathbf{x}\mid\mathbf{u}, s_k)$, normalised over the phase space at
plane $k$, setting the distribution of the survivors.

\paragraph{Transmission head.}
In our approach, the beam loss is treated via discrete-time survival analysis as a stochastic survival process where each configuration
launches $N$ ions and the simulation records the grouped survivor counts
$N \ge n_1 \ge \dots \ge n_{10}$~\citep{prentice1978grouped,tutz2016modeling}. The learnable hazard function $\lambda_k(\mathbf{u})$ is modeled as the probability that an ion admitted to plane $k-1$ is lost before plane $k$, and the transmission is the survival function, and the counts
follow a chain of
binomials whose trials are the survivors of the preceding plane,
\begin{equation}
\small
  T_k(\mathbf{u}) \;=\; \prod_{j=1}^{k}\bigl(1-\lambda_j(\mathbf{u})\bigr),
  \qquad
  n_k \mid n_{k-1} \sim
  \mathrm{Binomial}\!\bigl(n_{k-1},\, 1-\lambda_k(\mathbf{u})\bigr).
  \label{eq:survival}
\end{equation}
The hazards are read plane by plane from a latent beam state that a per-plane Gated Recurrent Unit (GRU)~\citep{cho2014gru} advances along the beam axis, in the spirit of neural discrete-time survival models~\citep{gensheimer2019scalable}.
The recurrence matches the optics, where the loss at a plane is set by the conditioning of the upstream electrodes, and the voltages are re-read
through a learned embedding. The
encoder, recurrence, and hazard function are detailed in
Appendix~\ref{app:transmission}.

\paragraph{Phase-space head.}
The kinetic ion distribution in a given plane is modeled as a normalizing
flow~\citep{papamakarios2021flows}. That is, an invertible map $f_\theta$ carries a
Gaussian base variable $\mathbf{z}\sim\mathcal{N}(\mathbf{0}, I_6)$ to the phase
space at any plane, $\mathbf{x} = f_\theta^{-1}(\mathbf{z}; \mathbf{u}, s_k)$, conditioned on the electrode voltages $\mathbf{u}$ and the arc length $s_k$. We implement this map as a rational-quadratic neural spline flow (RQ-NSF)~\citep{flow} composed of a stack
of coupling layers with tractable Jacobian determinants~\citep{realnvp}. Hence, it yields exact sampling together with an exact density in one pass through the
change of variables
$\log p_\theta(\mathbf{x}\mid \mathbf{u}, s_k)
= \log \mathcal{N}\!\bigl(f_\theta(\mathbf{x}; \mathbf{u}, s_k)\bigr)
+ \log\bigl|\det \partial f_\theta/\partial \mathbf{x}\bigr|.$

The flow is conditioned on the standardized voltages, using a fixed Fourier encoding~\citep{tancik2020fourier}
of the arc length $s_k$, so that a single flow represents all ten planes and we train it by
maximum likelihood using the surviving ions. The architecture and training are
detailed in Appendix~\ref{app:phasespace}.

\section*{Results}
We demonstrate the principle of our surrogate by training on a $60\%$ subset of the 65k simulated
configurations. The model accuracy is evaluated across the sampled population and beyond the transmission range seen during training, on settings refined during transmission optimization. These regimes define the population and high-transmission sets, with nearby configurations kept within the
same split. Their composition is given in Appendix~\ref{app:simion}.
We first examine how accurately the two heads recover the surviving mass and
phase-space distribution of the beam.

\begin{figure}[!t]
\centering
\includegraphics[width=0.9\textwidth]{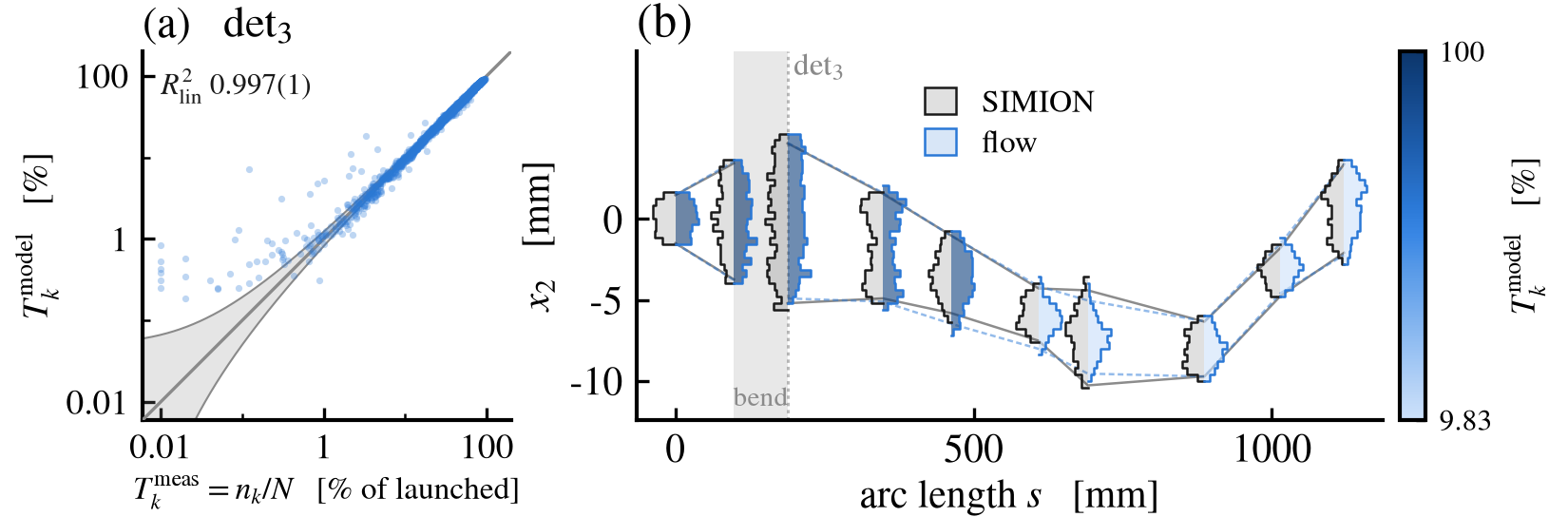}
\caption{\textbf{Surrogate predictions.} (a)~Predicted against measured
transmission at $\mathrm{det}_3$ over the delivering test configurations. The grey band is the 95\% binomial interval of the measurement for $10^4$ ions, and the
vertical columns at the left edge are its quantization in units of one delivered
ion. Metrics are means over three training seeds
with the sample standard deviation in parentheses. (b)~Transverse position $x_2$ at the ten detector planes for the
highest-transmission configuration of the dataset (test).
}
\label{fig:main-pair}
\end{figure}

\paragraph{Ion-survival.}
The first of the two heads recovers the surviving mass. It tracks the mass through the plane where the beam is lost by regressing the disappearance probability at each plane, and its $R^2$ stays above 0.94 at every detector. An example of this for $\mathrm{det}_3$ is shown in
Figure~\ref{fig:main-pair} a).

The measurement is a binomial proportion of $N$ launched ions, so counting variance sets a floor no model can beat, and the head sits a small factor above it across the predicted range (Figure~\ref{fig:transmission_floor}). The relative error therefore grows along the beamline as the counts thin, from a few per cent before the strongest aperture cut to roughly a third at the last detector (Figure~\ref{fig:transmission_arclength}), where the head still separates delivering from lost configurations with an AUC of 0.972(1).
However, training only on the delivering configurations removes every lost example, and the separation of delivering from lost settings then collapses at the planes before the aperture.
Further benchmarks and comparison with alternative transmission approaches are presented in Appendix~\ref{app:figures}.

\paragraph{Phase-space emulation.}

The phase-space head is evaluated on the same held-out configurations. Its samples track the SIMION phase space and reproduce its spread; the transverse $x_2$ marginal along the beamline for the highest-transmission held-out configuration is shown beside the transmission in Figure~\ref{fig:main-pair}. A quantitative comparison of several phase-space models across detector planes is presented in Appendix~\ref{app:figures}. At $\mathrm{det}_6$, immediately after the strongest cut in the geometry, the emulated emittance grows relative to SIMION and the centroid is offset by a third of the SIMION spread. The emittance ratio returns to unity by $\mathrm{det}_{10}$, however, the model can still constrain collimation at the end of the beamline. The kinetic-energy marginal degrades as the transmission falls; the Wasserstein distance to the SIMION marginal correlates with $\log T$ across held-out configuration–detector pairs (more details in the Appendix Figure~\ref{fig:energy-emd}).

We now turn to two applications of the differentiable surrogate. First, optimizing operating conditions by gradient ascent, and then inferring the beam in planes carrying no detector.
\paragraph{Surrogate-based constrained optimization.}

The transmission head is differentiable in the voltages, so $\log T_{10}$ is ascended directly through a bagged ensemble, whose members disagree away from the data and so keep the search near supported settings (Appendix~\ref{app:opt}).
The predicted maximum from gradient ascent achieved up to \(19.2\%\) when
validated in SIMION, above the \(16.7\%\) best setting found during the
black-box optimization campaign used in data generation (Figure~3a), and
above CMA-ES~\citep{hansen2001cma} run directly on SIMION from the same starts
(Appendix~\ref{app:opt}).

\begin{figure}[!t]
\centering
\includegraphics[width=0.88\textwidth]{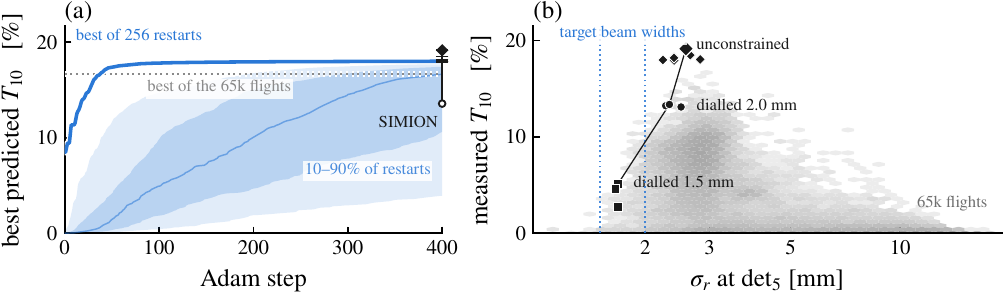}
\caption{\textbf{Gradient-based optimization through the surrogate.}
(a) Predicted $T_{10}$ during 256 gradient-ascent restarts, with selected
candidates validated in SIMION. The best candidate reaches 19.2\%
transmission, exceeding the best value in the original campaign (dotted).
(b) SIMION-validated optimization under beam-width constraints at
$\mathrm{det}_5$ showing the transmission--width trade-off, surrogate optima.}
\label{fig:optimization}
\end{figure}

After training, a direct SIMION evaluation traces $10^4$ ions per voltage setting in roughly $10\,\mathrm{s}$, whereas the surrogate evaluates and differentiates the final transmission in $\mathrm{ms}$. Batched over 256 restarts, each gradient step takes about $4\,\mathrm{ms}$, allowing hundreds of steps in seconds before a few candidates are re-flown in SIMION. Expensive particle tracking is thereby moved from the optimization loop to final validation. The differentiable flow also permits phase-space constraints beyond scalar transmission. We demonstrate this by imposing an augmented-Lagrangian equality on the predicted beam width at $\mathrm{det}_5$ (Appendix~\ref{app:opt}); SIMION validation shows that the delivered width decreases monotonically with the target (Figure~\ref{fig:optimization}b).

\paragraph{Inference of the unobserved beam.}
\label{sec:inference}

\begin{figure}[t]
\centering
\includegraphics[width=0.8\textwidth]{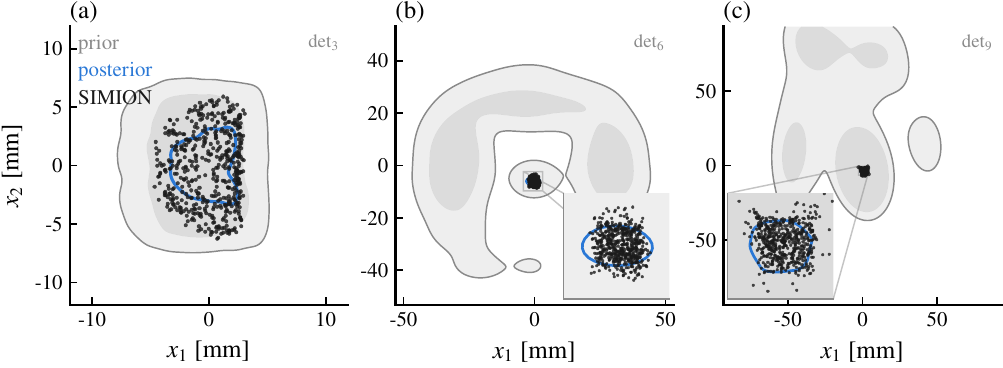}
\caption{\textbf{Inference of the beam shape.} Transverse
beam spot at three unobserved planes for the highest-transmission held-out
configuration. The grey
regions are the 68\% and 95\% highest-density regions of the prior predictive,
the beam the sampled voltage distribution admits at each plane, and the blue
region is the 68\% posterior predictive of Equation~\eqref{eq:pushforward},
conditioned on the counts and exit image of Equation~\eqref{eq:obslik} and
magnified in the insets where the prior dwarfs it; the SIMION ions are shown in
black.}
\label{fig:pushforward}
\end{figure}

We also demonstrate reconstruction at planes without detector observations. 
Let $\mathcal{O}=\{k_1<\dots<k_m\}$ index the instrumented planes, each reporting a count
$n_{k_i}$ and, where the detector images the beam, $M_{k_i}$ positions. Then, Eq. ~\eqref{eq:ansatz} turns  $\mathcal{D}=\{n_{k_i};\,\mathbf{y}^{(k_i)}_j\}$ into
\begin{equation}
\small
  p(\mathcal{D}\mid\mathbf{u}) \;=\;
  \prod_{i=1}^{m}
  \mathrm{Bin}\!\Bigl(n_{k_i};\, n_{k_{i-1}},\, \tfrac{T_{k_i}}{T_{k_{i-1}}}\Bigr)
  \prod_{j=1}^{M_{k_i}}
  p_\theta\bigl(\mathbf{y}^{(k_i)}_j \mid \mathbf{u},\, s_{k_i}\bigr),
  \label{eq:obslik}
\end{equation}
with $T_k \equiv T_k(\mathbf{u})$, $(n_{k_0}, T_{k_0}) = (N, 1)$, and
$p_\theta(\mathbf{y}|\mathbf{u},s_k)$ the flow marginal on the observed coordinates.
Each binomial is the survival between instrumented planes. To showcase this, we take $\mathcal{O}=\{4,7,10\}$ with
imaging at the exit plane only, where the marginal is evaluated with the
Gaussian $\mathcal{N}(\mathbf{y}_j;\mathbf{m}(\mathbf{u}),C(\mathbf{u}))$
(Appendix~\ref{app:inference}).
Since the posterior over $\mathbf{u}$ stays close to
the prior and the quantity of interest is extrapolated to an unobserved plane,
\begin{equation}
\small
  p(\mathbf{x}\mid s_k,\mathcal{D})
  \propto
    \int p_\theta(\mathbf{x}\mid\mathbf{u},s_k)\,
    T_k(\mathbf{u})\,q(\mathbf{u}\mid\mathcal{D})\,d\mathbf{u}
  .
  \label{eq:pushforward}
\end{equation}
This posterior predictive distribution describes the phase space of an ion conditional on its survival to plane $k$. We evaluate it by drawing voltage configurations from $q(\mathbf{u}\mid\mathcal{D})$, weighting them by their survival probabilities $T_k(\mathbf{u})$, and drawing ions from the conditional flow. Even when the voltages are uncertain, this pushforward concentrates onto the SIMION ions at the deep planes, shrinking the credible region while still covering them (Figure~\ref{fig:pushforward}). The contraction weakens at $\mathrm{det}_3$, and at the aperture plane the flow's own spread keeps the recovered spot broad (Appendix~\ref{app:inference}).

\section*{Conclusions}



We introduced a differentiable probabilistic surrogate for ion beam transport
in regimes dominated by particle loss. By factorizing the simulator response
into a survival probability and a conditional phase-space density, the model
combines discrete-time survival analysis with conditional normalizing flows to
provide a tractable likelihood for both ion counts and phase-space
observations, which we demonstrate by applying it to a high-dimensional electrostatic beamline.

Once trained, the surrogate replaces repeated particle-tracking simulations
with millisecond-scale differentiable evaluations and gradient calculations,
enabling direct gradient-based optimization even past the optimum found with
traditional ion-beam optimization approaches.
The same likelihood also enables optimization under beam-shape constraints and inference of the ion beam at unobserved locations from sparse detector measurements.

The main remaining challenge is to close the gap between simulation and
experiment. The largest errors in the predicted beam shape occur after strong
geometric cuts, where particle loss strongly shapes the distribution of ions
that survive. Future work should link the survival and beam-shape models more
closely and calibrate the surrogate with experimental data. Such calibration
could account for imperfect field maps, electrode misalignments, changes in
source conditions, environmental drift, and detector effects. Combined with
optimization that accounts for uncertainty, this would turn the present
simulation-based surrogate into a digital twin for precision experiments with
rare isotopes.

This approach is currently being extended to nuclear precision experiments at MIT and Texas A\&M University, which serve as testbeds for an ongoing DOE Genesis Mission project, where the surrogate is trained on data from a GPU-accelerated, differentiable particle tracker and used for beamline optimization.

\section*{Acknowledgments}

We acknowledge support from the PCLB Foundation and the U.S. Department of Energy, Office of Science, under Award Numbers  DE-SC0026800 and DE-SC0026653 (Genesis Mission). Computational resources were provided by SubMIT at MIT Physics.
We thank Scott Moroch, Alex Brinston, and Arian Jadbabaie (MIT), Jonas Karthein (Texas A\&M University), and the DOE Genesis FastSims project for insightful discussions.

\subsubsection*{Data Availability}

The code and data needed to reproduce the results reported in this
paper are publicly available on GitHub~\cite{Munoz2026IonBeamSurrogate}.

{\small
\bibliography{references}
}

\appendix


\newpage

\section{Appendix: Architecture details and Training}
\subsection{Transmission head architecture and training}
\label{app:transmission}
The transmission head predicts, for a voltage configuration $\mathbf{u}$, the
conditional survival probability $p_k(\mathbf{u})$ that an ion reaching plane
$k-1$ is delivered to plane $k$, and the transmission follows as the running
product $T_k = \prod_{j\le k} p_j$ of Equation~\eqref{eq:survival}. The beam
crosses the planes in order, so the head is a recurrence unrolled along the
beamline over the eight planes after the first bend, $\mathrm{det}_3$ through
$\mathrm{det}_{10}$. The first two planes are pinned to $T=1$, because every
configuration in the dataset delivers all $10^4$ launched ions to them.

A forward pass reads the $22$ standardised electrode voltages and carries a
hidden state of width $48$ down the line. Each plane sees the voltages through a
soft gate $\mathbf{v}_k = \mathbf{u}\odot\mathrm{sigmoid}(\mathbf{g}_k)$, whose
logits $\mathbf{g}_k$ start at $+1$ on the electrodes grouped with plane $k$ and
$-1$ elsewhere, so the cell attends mainly to the electrodes that steer the beam
through it, and are trained with the rest of the model. A two-layer encoder, a
linear map to width $48$ followed by a SiLU nonlinearity and a second linear map,
turns the full voltage vector into the initial state
$\mathbf{h}_0\in\mathbb{R}^{48}$, the beam state before any transport. A gated
recurrent cell then advances the state one plane at a time,
$\mathbf{h}_k = \mathrm{GRU}(\mathrm{inp}_k(\mathbf{v}_k)+\mathbf{e}_k,\,
\mathbf{h}_{k-1})$, with a per-plane input projection $\mathrm{inp}_k$ and a
trainable per-plane embedding $\mathbf{e}_k$, and a linear head with a per-plane
intercept reads the hazard logit whose sigmoid is $p_k$. The cells and heads are
untied across planes, so each plane learns its own aperture geometry, which brings
the head to about $1.3\times10^5$ parameters.

Training maximises the likelihood of the observed survivor counts over every
configuration, the failing ones included. Because each plane is weighted by the
ions that reach it, a lost setting still constrains the planes it does populate,
and the head learns to tell delivering configurations from lost ones. The fit
term is the trial-weighted binomial negative log-likelihood, equal to a binary
cross-entropy on the per-plane survival fraction $n_k/m_k$ with trials
$m_k = n_{k-1}$ and weight $m_k$, so a plane that receives no ions carries no
weight. A mean-squared error on $\log_{10}T_{10}$ at weight $0.1$, over the
configurations that deliver ions to every plane, is added to balance the fit
across the transmission range without disturbing the ranking. Optimization uses
Adam~\citep{kingma2015adam} with a coupled weight decay of $10^{-5}$ at a learning
rate of $2\times10^{-3}$, batch size $1024$, cosine annealing, gradient-norm
clipping at $1.0$, and early stopping on the validation objective.

A point prediction is paired with a predictive law for the survivor counts. The
committed law adds a per-plane dispersion to the survival chain. A beta-binomial
head models the count as
$n_k \mid n_{k-1} \sim \mathrm{BetaBinomial}(n_{k-1}, p_k\phi_k, (1-p_k)\phi_k)$,
inflating the binomial variance by a factor $\phi_k(\mathbf{u})$ that returns the
binomial as $\phi_k\to\infty$. It is fit as a second stage on the frozen hazard
mean, so it changes no point prediction and supplies only the interval, with the
survival factors $p_k$ exactly those of the committed head. Pooled over the eight
planes it covers $0.794(2)$ of held-out counts at a nominal $68\%$ and $0.959(1)$
at $95\%$, against $0.391$ and $0.532$ for the plain binomial, calibrated at the
upper level and a little conservative at the lower, and it is the interval drawn
in Figure~\ref{fig:parity-si}. A dispersion trained jointly with the mean comes
out optimistic, by a factor of two to thirteen across the planes, which is why the
fit is staged on the frozen residual.

\subsection{Phase-space head architecture and training}
\label{app:phasespace}
The phase-space head is a conditional normalising flow that maps the six
standardised phase-space coordinates of a single ion to a standard Gaussian in
$\mathbb{R}^6$ and back, and so returns the per-ion density in closed form. It is
a rational-quadratic neural-spline coupling flow of six coupling layers whose
active-coordinate masks cycle, so that each coordinate is transformed several
times and none is left untouched. A coupling layer holds three coordinates fixed,
passes them together with the conditioning context through a network of two hidden
layers of width $256$, and shapes from its output the monotone rational-quadratic
splines of eight bins on $[-5,5]$, with identity tails outside, through which the
other three coordinates are mapped. The coordinates are per-plane standardised, so
the interval covers the bulk of every plane, and composing the layers gives an
exact, invertible change of variables from the ion to the base Gaussian.

Every layer draws on a shared context that encodes the operating point. A
conditioning network combines a $64$-dimensional embedding of the standardised
voltages, gated per plane by a trainable electrode mask, with sixteen geometric
Fourier frequencies of the normalised arc length and eight of the normalised log
noise scale, each read out in sine and cosine, into the $128$-dimensional context
supplied to every layer. The arc-length conditioning is trained and evaluated only
at the ten measured planes, so densities between planes are not validated. The
gate is a per-plane table of electrode weights read through a sigmoid, initialised
to damp the electrodes that follow the plane in electrode-index order; the prior
is weak, because that order differs from the beam order, and shallow enough that a
few hundred training steps override it. The flow holds about $7.5\times10^{5}$
parameters.

Training maximises the exact per-ion log-likelihood of the ions, each perturbed by
Gaussian noise of a scale that is itself supplied to the flow. Planes are drawn
uniformly and plane $k$ is weighted by $\sqrt{\max_j n_j / n_k}$ normalised to
unit mean, which lifts the ion-starved deep detectors by about four relative to
the first plane, and the store holds at most $512$ ions per configuration and
plane. The noise scale is drawn log-uniformly over $[0.01, 0.5]$ in standardised
units during training and fixed to $0.01$ at evaluation. Optimization uses AdamW
at a learning rate of $3\times10^{-3}$, weight decay $10^{-5}$, global-norm
gradient clipping at $5$, and a warmup-cosine schedule, over batches of $64$
configurations at $64$ ions each.

The noise conditioning follows noise-conditioned score-based and diffusion
models~\citep{song2019score,ho2020ddpm,karras2022edm}. Perturbing the ions by
$\sigma\boldsymbol{\varepsilon}$ with
$\boldsymbol{\varepsilon}\sim\mathcal{N}(\mathbf{0}, I_6)$ and conditioning on
$\sigma$ makes the flow represent a one-parameter family of densities
$p_\theta(\mathbf{x}\mid \mathbf{u}, s_k, \sigma)$, each the exact law of the clean
phase space convolved with $\mathcal{N}(\mathbf{0}, \sigma^2 I_6)$. At the evaluation
value $\sigma = 0.01$ the reported density is the phase space smoothed by a variance
of order $10^{-4}$ against unit-variance coordinates, a smoothing whose magnitude is
known; densities are comparable across models only at equal
$\sigma$. A sweep of $\sigma$ below the training floor of $0.01$ raises the held-out
negative log-likelihood again, which is extrapolation in the conditioning variable,
 so we report at the floor. Conditioning on the noise scale keeps the
likelihood smooth and everywhere supported, so that gradient ascent and Markov-chain
inference over the
voltages~\citep{papamakarios2019snl,albergo2019flowmcmc,noe2019boltzmann} are not
obstructed by the sharp, near-degenerate densities a coupling flow would otherwise
place on the ion distribution.

\section{Appendix: SIMION simulation and dataset}
\label{app:simion}

The simulation covers an electrostatic transport line that carries molecular
ions from an ion source to the injection point of a cryogenic buffer-gas
cell. The simulated line holds 42 electrodes; 22 of their voltages are free, the
source electrode is pinned at $500\,$V, and the rest sit at ground. The free
electrodes group into an extractor, two einzel lenses, an injection tube, four
front and four back steerers, a quadrupole bender with four poles and two lenses,
and four exit steerers, where the bender turns the beam through $90^\circ$
between the second and third detector plane. The hardware ranges span from
$-1300\,$V at the extractor to $\pm1200\,$V at the bender poles, with steerers
within $150\,$V; one standardised voltage unit corresponds on average to
$61\,$V per electrode over the training distribution. The potential arrays and the ray
tracing are computed with SIMION~8.2~\citep{dahl2000simion}. Each flight launches
singly charged ions of mass $156\,$amu from a uniformly filled disc of radius
$3.3\,$mm, with initial directions filling a cone of half-angle $40^\circ$ about
the beam axis and a Gaussian kinetic energy of mean $0.25\,$eV and FWHM
$10^{-4}\,$eV.

The ten detector planes are virtual crossing planes at arc lengths $s_k = 0$,
$98.1$, $188.1$, $347.4$, $462.8$, $608.9$, $691.5$, $887.0$, $1014.1$, and
$1120.9\,$mm, spaced non-uniformly by the hardware layout, with the first two
planes normal to the injection axis and the remaining eight normal to the exit
axis after the bend. Each crossing records the six phase-space coordinates
defined in the Methods, the positions in mm, the direction cosines
dimensionless, and the kinetic energy in eV. Three of the ten planes carry
detectors in the physical apparatus, the two Faraday cups at $s_4$ and
$s_7$ and the exit plane $s_{10}$ adjacent to the channeltron; the remaining
planes are observable only in simulation.

Uniform sampling of the hardware ranges almost never delivers beam to the deep
planes, because the settings that transport at all fill a productive region only
about $10^{-7}$ of the hardware voltage volume. The dataset is therefore assembled
from four complementary sampling strategies. A Latin Hypercube sampling exploits the
productive region quasi-uniformly. Gaussian jitter around settings already known
to reach the deep planes, at a scale of about a tenth of each electrode's
productive span, surrounds each of them with many nearby settings, so those planes
are seen under varied conditions. A separate transmission-optimization campaign
contributes its ten thousand highest-$T_{10}$ settings. A final acquisition batch
places settings on a shell around the known high-transmission region and keeps
those a gradient-boosted model of delivery and transmission scores well, so the
scarce high-yield settings are oversampled rather than left to chance. The last
two strategies exist to populate the high-transmission regime that the first two
barely reach.

The flights ran as a SLURM array of one-hour chunks on four CPU cores each, at
about a second of tracing per thousand ions, with the runner skipping any
configuration already computed so that the campaign of $6.5\times10^{8}$
trajectories is resumable. Under broad, unguided sampling the mean transmission
falls from unity at the first two planes to about $10^{-3}$ at the last, where the
final binomial of Equation~\eqref{eq:survival} then carries only some twenty ions
of signal.

The settings divide into a training set of $38{,}707$, a validation set and a
test set of $8{,}294$ each, and two further sets held out whole for the
evaluations the main text reports. Because the harvested and acquisition settings
cluster tightly, nearly every one with a near-duplicate within $0.5$ standardised
voltage units, the division is made over groups of near-duplicates rather than
single settings, so no test setting is a near-copy of one seen in training. The
high-transmission block, the $9{,}301$ settings in the top $1\%$ by measured
$T_{10}$, is held out to probe extrapolation past the training ceiling of
$10.6\%$, up to $16.7\%$ at the dataset best. The campaign block of $404$ settings
is held out whole from the optimization campaign and supplies its baseline in the
main-text comparison. The population block, $5{,}729$ settings, is the subsample
of the test set whose delivery rate matches the natural population, $37\%$ of
settings reaching the last plane, and it carries the emulation metrics.

Training one transmission-head seed takes $\sim$5 minutes and one flow seed
$\sim$20 minutes on a single GTX~1080~Ti, the GPU used for every timing quoted
in the main text. The training and evaluation code, the trained weights, and
the 65k dataset
are available from the authors upon request.

\section{Appendix: Inference procedure}
\label{app:inference}

The observation model of Equation~\eqref{eq:obslik} treats survival as Markov
along the beam, so each observed count is the trial size of the next binomial in
the chain of Equation~\eqref{eq:survival}. The exit image enters through a
plug-in synthetic likelihood in its first two moments, and
reparameterised flow samples under common random numbers keep
$\mathbf{m}(\mathbf{u})$, $C(\mathbf{u})$, and the likelihood itself smooth in
$\mathbf{u}$.

Under a standard-normal prior on the standardised voltages, the posterior
$q(\mathbf{u}\mid\mathcal{D})$ is a Laplace approximation at the gradient-ascended
optimum, its Hessian differentiated exactly through both heads, and the few
directions of weak or locally non-concave curvature are clamped to the prior
scale. The pushforward of Equation~\eqref{eq:pushforward} is
evaluated by sampling voltages from $q$ and ions from the flow, over the held-out
settings bright enough to image at the exit plane, and the contraction reported in
the main text is a median over those settings.

The contraction is largest at the deep planes and weakest at $\mathrm{det}_3$,
upstream of the first cup, where prior and posterior share the same scale and the
posterior turns overconfident. At the aperture plane the recovered spot stays
several times broader than the SIMION one, the diffuseness the flow itself carries
there. Conditioning instead on complete six-dimensional phase-space clouds, an
observation no apparatus provides, tightens the voltage posterior further but
leaves it mis-centred, because at that statistical power the residual
emulator--SIMION mismatch dominates the inference. Conditioning at the fidelity
the hardware delivers keeps the recovered beam consistent with SIMION.

Replacing the flow by the conditional Gaussian of Table~\ref{tab:phase_space}, in
both the likelihood and the pushforward, tests what the flow brings to the
inference. Over 65 held-out settings, the median ratio of reconstructed to SIMION
spot width at $\mathrm{det}_3$, $\mathrm{det}_5$, $\mathrm{det}_8$, and
$\mathrm{det}_9$ is 1.3 to 2.5 for the Gaussian, with centroid offsets of 0.5 to
1.8 SIMION spreads, against 0.9 to 1.3 and 0.2 to 0.7 for the flow. With
observations sampled from the Gaussian model, the same inference recovers the
widths within 20\%, so the discrepancy originates in the Gaussian beam shape.

\section{Appendix: Optimization procedure}
\label{app:opt}

\paragraph{Which model produces which number.} The accuracy of the hazard head
itself is measured for a single head over three independently trained seeds, in
the benchmark of Table~\ref{tab:transmission}, its distance from the shot-noise
floor, and its error along the beamline and across the measurement range. The
committed model, deployed for the calibrated parity of the main text with its
per-plane repeat, the extrapolation to the highest-transmission settings, and the
optimization, is the twelve-member ensemble, each member trained on its own
bootstrap resample of the training rows and its own initialisation and combined as
the mean of $T_k$ across members. The two are not interchangeable out of distribution. Ascending a single head
drives the voltages to $3.8$ to
$4.0$ standardised units from the nearest training configuration, where it
predicts near 23\% and SIMION measures 13.2 to 16.6\%, an overprediction of six
points. The ensemble stops at $1.7$ to $2.5$ units and predicts within about
half a point of what SIMION then measures. The mechanism is disagreement among
the members. A member trained without a given region of voltage space scores it
low, the mean
follows, and the ascent is held where the members agree, which is where the
data is. Selecting endpoints by a lower confidence bound, mean minus the member
spread, or inside an explicit trust region on the distance to the data changes
nothing measurable once the ascent itself runs through the ensemble; the
robustness is a property of the averaged surface.

Both ascents run over unconstrained variables carried into the training voltage
box by a tanh reparameterisation,
$\mathbf{u} = \mathbf{l} + \tfrac{1}{2}(\mathbf{h}-\mathbf{l})(\tanh\mathbf{w}+1)$
per electrode in standardised units, with $[\mathbf{l}, \mathbf{h}]$ the
per-electrode range of the training split, so every iterate is a settable
configuration. Half the restarts start at measured configurations drawn from the
training and natural test rows and half uniformly in the box, both pulled away
from the boundary before the inverse map. The unconstrained ascent runs 256
restarts for 400 Adam steps at a learning rate of 0.02; the gradient of
$\log T_{10}$ with respect to the voltages costs 0.34\,ms for a single
configuration on the GTX~1080~Ti; batched over the restarts, about 4\,ms per
step.

The constrained variant reads the RMS transverse width at $\mathrm{det}_5$ off
the flow, $\sigma_r = (\operatorname{var} x_1 + \operatorname{var} x_2)^{1/2}$
in physical millimetres over reparameterised samples
$\mathbf{x} = f_\theta^{-1}(\mathbf{z}; \mathbf{u}, s_5)$, with 96 base draws
$\mathbf{z}$ fixed across steps and restarts so the read is a smooth
deterministic function of the voltages. The target enters as an augmented
Lagrangian, the objective $\log T_{10} - \lambda c - \tfrac{\rho}{2}c^2$ with
$c = \sigma_r - w^\ast$, a per-restart signed multiplier $\lambda$ updated every
step by dual ascent at 0.1 per mm of violation and clipped to $\pm 50$, and
$\rho = 5\,$mm$^{-2}$; 128 restarts of 400 steps complete in two minutes on a
single CPU. Endpoints are scored on 1{,}024 independent draws and count as
feasible within 5\% of the target. The constrained ascent is run at two dialled
widths, 2.0 and 1.5\,mm, against the 2.7\,mm the flow reads at the
unconstrained optima; the tighter dial leaves 24 of the 128 restarts feasible
against 42 at the looser one.

Verification re-flies the selected settings from scratch through the production
SIMION runner (\texttt{prod\_evaluate}) at $10^4$ ions per setting, the flight
protocol of the dataset itself; the unconstrained ascent contributed six best
restarts and one drawn deliberately from the far tail of the distance to the
training data, and each constrained ascent its three best feasible ones, in
every case thinned to a mutual separation of at least one standardised unit,
for thirteen re-flown settings in all. The
re-flown $\mathrm{det}_5$ clouds are projected into the frozen plane frames of
the dataset before the width is taken, so the measured widths are commensurate
with the flow's read.

\paragraph{Constraint after the aperture.} The same ascent is run with the
constraints placed at the exit plane, $|\bar{x}_{1}|, |\bar{x}_{2}| \le 1\,$mm on
the centroid and $\sigma_{x_1} \le 1.09$, $\sigma_{x_2} \le 1.28\,$mm on the
widths, the median spot of the high-transmission block; 43 of the 256 restarts
end feasible. The three best feasible settings, re-flown as above, deliver 9.5,
9.0, and 6.3\% against the predicted 9.9, 7.9, and 6.2\%, with SIMION widths of
1.0 to 1.5\,mm and centroids within 1\,mm in two of the three.

\paragraph{Black-box baseline.} As a reference, CMA-ES~\citep{hansen2001cma} is
run directly on SIMION from the starting settings of the seven re-flown restarts
of Figure~\ref{fig:optimization}a, within the same voltage box and with $10^4$
ions per flight. The objective is
$T_{10} + 10^{-3}\sum_{k=1}^{10} 4^{k}T_k \big/ \sum_{k=1}^{10} 4^{k}$, which
remains informative where no ion reaches the last plane. After $10^3$ flights per
start, its best settings reach 3.2 to 17.5\% when re-flown, below the 13.6 to
19.2\% of the gradient-ascent endpoints from the same starts, in all seven cases.

\section{Appendix: Emulator accuracy}
\label{app:figures}

Transmission accuracy is a question of counts, and phase-space accuracy a question
of distributions. The two heads are evaluated separately against SIMION on the
held-out blocks of Appendix~\ref{app:simion}, each result placed beside the
sampling floor it is measured against.

\subsection{Transmission head}
\begin{figure}[t]
\centering
\includegraphics[width=0.6\textwidth]{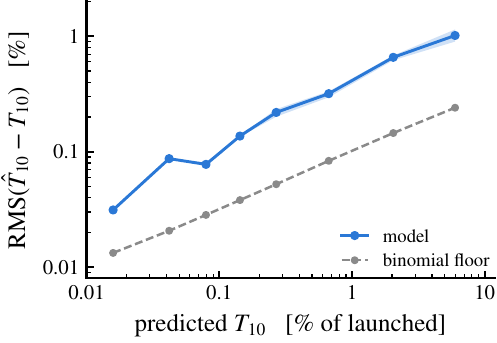}
\caption{\textbf{Model error against the shot-noise floor.} RMS residual of the
predicted $T_{10}$ in octile bins of the prediction on
the population block, the mean over three seeds with the band spanning the
seed range, beside the binomial error a model returning the exact
transmission would still show. The two curves run parallel across nearly four
decades; the model sits a factor 2.4 to 4.5 above the floor (4.2 pooled).}
\label{fig:transmission_floor}
\end{figure}

\begin{figure}[t]
\centering
\includegraphics[width=0.6\textwidth]{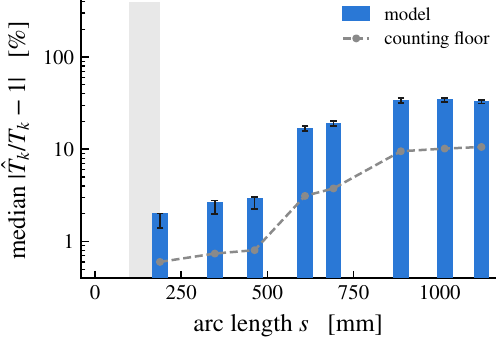}
\caption{\textbf{Error along the beamline.} Median relative error
$|T_{\mathrm{model}}/T_{\mathrm{meas}}-1|$ over the delivering
configurations of the population block, at each plane, placed at the
plane's arc length; bars span the three training seeds, and the grey curve is
the same statistic under pure binomial counting noise, obtained by Monte
Carlo. The shaded span marks the bend between $\mathrm{det}_2$ and
$\mathrm{det}_3$; the excess over the shot-noise floor is a factor 2.7 to 5.3,
largest at $\mathrm{det}_6$ and $\mathrm{det}_7$, the planes just after the strongest
aperture cut.}
\label{fig:transmission_arclength}
\end{figure}

\begin{figure}[t]
\centering
\includegraphics[width=0.9\textwidth]{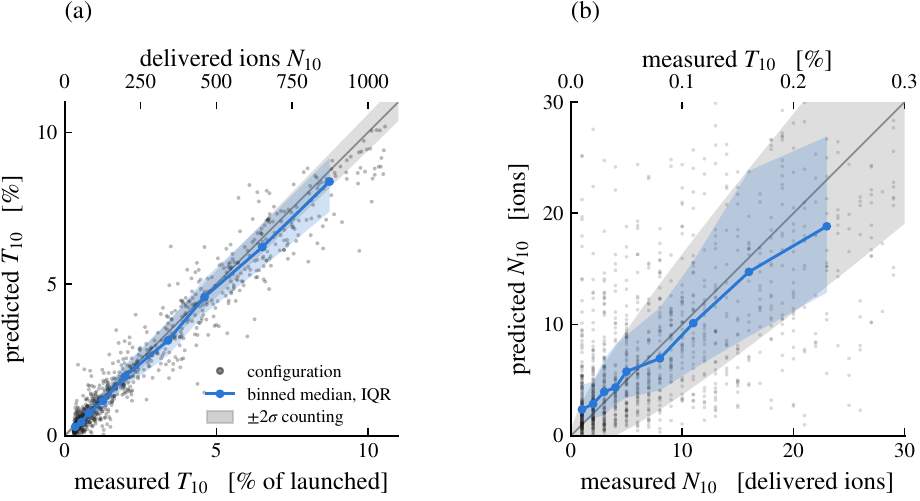}
\caption{\textbf{Transmission parity across the measurement range.} Predicted
against measured $T_{10}$ over the delivering configurations of the population
block, the single head at seed~0 with the binned median and interquartile band
taken over three seeds. The set is split at thirty delivered ions into (a)~the
well-measured configurations, shown in transmission, and (b)~the rest, shown in
delivered ion counts where the measurement is itself only a handful of ions. The
grey band is the central 95\% binomial counting interval about the identity, the
spread an exact model would still show against the finite ion count; the
prediction stays inside it down to the single-ion floor.}
\label{fig:transmission_parity}
\end{figure}

\begin{figure}[t]
\centering
\includegraphics[width=0.75\textwidth]{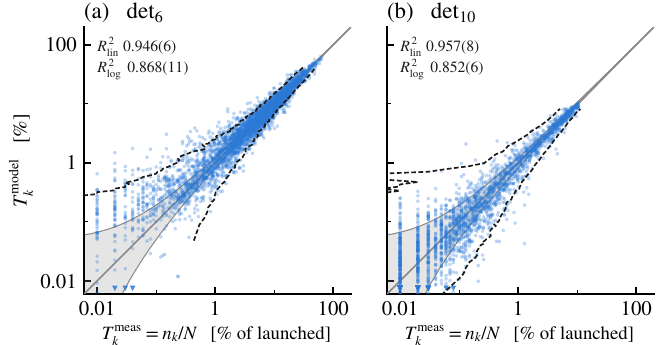}
\caption{\textbf{Transmission parity after the aperture.} Predicted against
measured transmission over the delivering configurations of the population
block: (a)~at $\mathrm{det}_6$, immediately after the strongest aperture cut, and
(b)~at $\mathrm{det}_{10}$, the last detector. Axes and annotations follow
Figure~\ref{fig:main-pair}(a), so all three planes sit on one scale. The grey
ribbon is the central 95\% binomial interval of the measurement and the dashed
lines are the model's own central 95\% interval from the beta-binomial dispersion
head of Appendix~\ref{app:transmission}; the vertical columns at the left edge
are the quantisation of the measurement in units of one delivered ion.}
\label{fig:parity-si}
\end{figure}

\begin{figure}[t]
\centering
\includegraphics[width=0.62\textwidth]{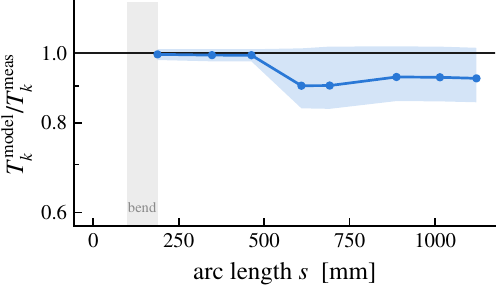}
\caption{\textbf{Extrapolation to the highest-transmission settings.} The ratio
$T_{\mathrm{model}}/T_{\mathrm{meas}}$ of the committed twelve-member ensemble to
the measurement over the $9{,}301$ configurations of the high-transmission block,
held out beyond the training range, at each plane against its arc length; the line
is the median over configurations and the band the $10$--$90\%$ range. The ratio
holds at unity through $\mathrm{det}_5$ and settles near $0.9$ beyond the aperture,
a systematic underprediction of the surviving mass at the frontier that stays
within about a tenth in the median.}
\label{fig:transmission_evolution}
\end{figure}

The head is benchmarked in Table~\ref{tab:transmission} against three baselines
that share its per-plane survival factorisation, a generalised linear model, a
gradient-boosted tree, and a multilayer perceptron. Delivery, the separation of
the configurations that reach the last plane from those that do not, and
magnitude, the accuracy of the surviving transmission, are both led by the hazard
head. Its error tracks the binomial shot-noise floor of the measurement to within
a small constant factor along the whole beamline
(Figures~\ref{fig:transmission_floor} and~\ref{fig:transmission_arclength}), and
the predicted counts fall inside the counting interval plane by plane
(Figures~\ref{fig:transmission_parity} and~\ref{fig:parity-si}). Extrapolation to
the highest-transmission settings, the one regime beyond the training range, is
where the committed twelve-member ensemble earns its place
(Figure~\ref{fig:transmission_evolution}).

\begin{table}[t]
\centering
\caption{\textbf{Transmission benchmark.} Delivery AUC, the separation of
delivering from lost configurations, over all configurations,
linear $R^2$ and RMSE over the delivering configurations, and Spearman $\rho$
over the population block, with RMSE and $\rho$ on the
high-transmission block; $R^2$ is not reported there, because
selection on a noisy measurement caps it near 0.25 for any model. The GLM,
GBDT, and MLP share the per-plane survival factorisation of
Equation~\eqref{eq:survival}. Metrics are means over three training seeds
with the sample standard
deviation in parentheses; the GLM is deterministic across seeds.}
\label{tab:transmission}
\vspace{0.5em}
\small
\begin{tabular}{lcccccc}
\toprule
 & \multicolumn{4}{c}{Population block} & \multicolumn{2}{c}{High-transmission block} \\
\cmidrule(lr){2-5}\cmidrule(lr){6-7}
Model & AUC & $R^2$ & RMSE [pp] & $\rho$ & RMSE [pp] & $\rho$ \\
\midrule
GRU hazard & 0.972(1) & 0.957(8) & 0.45(4) & 0.830(2) & 1.05(11) & 0.934(25) \\
MLP        & 0.933(2) & 0.828(6) & 0.91(2) & 0.769(3) & 2.86(23) & 0.702(5)  \\
GBDT       & 0.874(2) & 0.434(5) & 1.65(1) & 0.676(4) & 5.42(2)  & 0.368(13) \\
GLM        & 0.691    & 0.006    & 2.19    & 0.376    & 7.24     & $-0.267$  \\
\bottomrule
\end{tabular}
\end{table}

\subsection{Phase-space head}
\begin{figure}[t]
\centering
\includegraphics[width=0.62\textwidth]{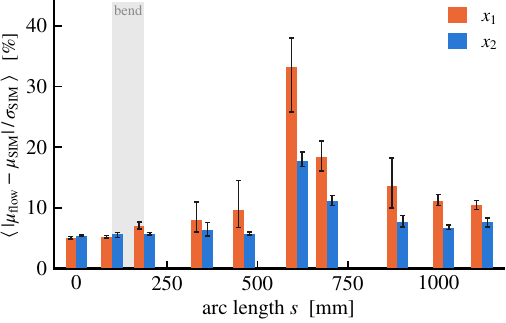}
\caption{\textbf{Centroid placement along the beamline.} Mean over 400 held-out
configurations of the centroid offset
$|\mu_{\mathrm{flow}}-\mu_{\mathrm{SIM}}|/\sigma_{\mathrm{SIM}}$, as a percentage
of the SIMION spread, for the two transverse positions $x_1$ and $x_2$ at each
plane, the configurations a random sample of the settings with at least 128 ions
at every plane and the error bars spanning the three training seeds. The shaded
span marks the bend between $\mathrm{det}_2$ and $\mathrm{det}_3$; the offsets
peak at $\mathrm{det}_6$ and $\mathrm{det}_7$, immediately after the strongest
aperture cut.}
\label{fig:flow-evolution}
\end{figure}

\begin{figure}[t]
\centering
\includegraphics[width=\textwidth]{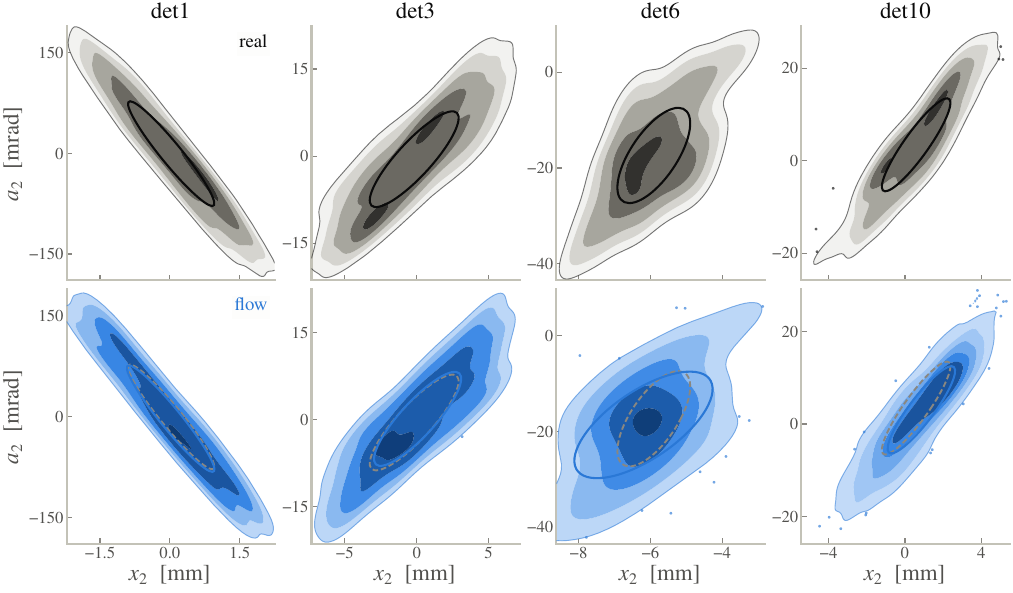}
\caption{\textbf{Conjugate phase space at four planes.} The $(x_2, a_2)$ projection for
SIMION ions (top) and flow samples at matched sample size (bottom), for the
configuration of Figure~\ref{fig:main-pair} (the dataset-best setting);
kernel-density contour levels are
computed on the SIMION panel and reused below it, so equal shade is equal density.
Solid ellipses mark the $1\sigma$ ellipse of each panel's own cloud, and the dashed
ellipse repeats the SIMION one on the flow panel.
Emittance ratios of flow to SIMION in the drawn projection, with bootstrap 95\%
intervals, read 1.01~[0.93, 1.09] at $\mathrm{det}_1$, 0.98~[0.90, 1.08] at
$\mathrm{det}_3$, 2.12~[1.37, 2.97] at $\mathrm{det}_6$, the plane after the strongest
aperture cut (population band across seeds 2.2 to 3.9), and 0.96~[0.86, 1.08] at
$\mathrm{det}_{10}$. The undrawn projection
reads 0.76~[0.64, 0.91] at $\mathrm{det}_3$, a single-configuration departure from a
population band of 1.03 to 1.15, and covers unity elsewhere away from
$\mathrm{det}_6$.}
\label{fig:phase-space}
\end{figure}

\begin{figure}[t]
\centering
\includegraphics[width=0.7\textwidth]{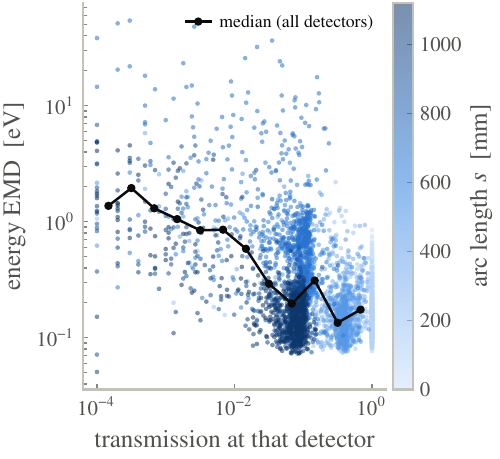}
\caption{\textbf{Energy-marginal error against transmission.} Each point is one
held-out (configuration, detector) pair: the Wasserstein distance between the SIMION and
emulated kinetic-energy marginals at a plane, against the transmission to that plane,
for up to 400 configurations per detector (4{,}000 points, seed 0), coloured by arc
length. The black line is the median over twelve log-spaced transmission bins and spans
$0.13$ to $1.95\,$eV; Spearman $\rho = -0.34$ between $\log T$ and the log distance
($p < 10^{-100}$).}
\label{fig:energy-emd}
\end{figure}

The flow is scored against SIMION at the exit plane by energy distance, a
distribution-level two-sample statistic, and by exact per-ion likelihood, in
Table~\ref{tab:phase_space}. Under energy distance the noise-conditioned flow
lands far closer to SIMION than the conditional-Gaussian and control baselines,
and the gap widens on the two extrapolation blocks; noise conditioning costs
nothing here, the conditioned and unconditioned flows agreeing within a seed
standard deviation on every block. The conditional Gaussian records the higher
per-ion likelihood, because the sharp flow density assigns near-zero mass to ions
on the sparsely covered strata while the Gaussian keeps a covariance broad enough
to cover them, and the energy distance rejects that over-spread density at a
$z$-score of tens. The centroid placement, the full conjugate phase space, and the
kinetic-energy marginal along the beamline appear in
Figures~\ref{fig:flow-evolution}, \ref{fig:phase-space}, and~\ref{fig:energy-emd}.

\begin{table}[t]
\centering
\caption{\textbf{Phase-space benchmark.} Energy-distance $z$-score against a
matched-size resampling null (lower is better, $\downarrow$) and per-ion
log-likelihood (higher is better, $\uparrow$) at the exit plane
$\mathrm{det}_{10}$, over the population, high-transmission, and campaign
blocks; the control carries no density, so its log-likelihood is undefined.
Values are means over three training seeds with the sample standard deviation
in parentheses.}
\label{tab:phase_space}
\vspace{0.5em}
\small
\setlength{\tabcolsep}{4.5pt}
\begin{tabular}{lcccccc}
\toprule
 & \multicolumn{3}{c}{Energy-distance $z$ ($\downarrow$)} & \multicolumn{3}{c}{Per-ion $\log p$ [nats] ($\uparrow$)} \\
\cmidrule(lr){2-4}\cmidrule(lr){5-7}
Model & Pop. & High-$T$ & Camp. & Pop. & High-$T$ & Camp. \\
\midrule
Flow                 & 4.9(2.1)  & 3.5(1.6)  & 1.8(0.5)  & $-4.77(2.64)$ & $-1.27(0.05)$ & $-1.12(0.02)$ \\
Flow, no noise cond. & 4.6(1.3)  & 4.0(2.4)  & 1.5(0.2)  & $-4.47(2.65)$ & $-1.14(0.07)$ & $-0.96(0.03)$ \\
Conditional Gaussian & 36.5(5.4) & 58.4(9.1) & 28.4(3.8) & $-2.02(0.15)$ & $-0.65(0.12)$ & $-0.76(0.13)$ \\
Control              & 45.3(7.1) & 54.5(8.8) & 45.5(7.6) & --            & --            & --            \\
\bottomrule
\end{tabular}
\end{table}

\end{document}